\documentclass[aps,prd,twocolumn,a4paper,superscriptaddress]{revtex4-1}
\usepackage{graphicx}
\begin{document}

\newcommand{\be}{\begin{equation}}
\newcommand{\ee}{\end{equation}}
\newcommand{\bq}{\begin{eqnarray}}
\newcommand{\eq}{\end{eqnarray}}
\newcommand{\bsq}{\begin{subequations}}
\newcommand{\esq}{\end{subequations}}
\newcommand{\bc}{\begin{center}}
\newcommand{\ec}{\end{center}}

\title{Cosmological and astrophysical constraints on tachyon dark energy models}

\author{C. J. A. P. Martins}
\email{Carlos.Martins@astro.up.pt}
\affiliation{Centro de Astrof\'{\i}sica da Universidade do Porto, Rua das Estrelas, 4150-762 Porto, Portugal}
\affiliation{Instituto de Astrof\'{\i}sica e Ci\^encias do Espa\c co, CAUP, Rua das Estrelas, 4150-762 Porto, Portugal}
\author{F. M. O. Moucherek}
\email[]{Fernando.Moucherek@astro.up.pt}
\affiliation{Instituto de Astrof\'{\i}sica e Ci\^encias do Espa\c co, CAUP, Rua das Estrelas, 4150-762 Porto, Portugal}
\affiliation{Faculdade de Ci\^encias, Universidade do Porto, Rua do Campo Alegre 687, 4169-007 Porto, Portugal}
\date{24 April 2016}

\begin{abstract}
Rolling tachyon field models are among the candidates suggested as explanations for the recent acceleration of the Universe. In these models the field is expected to interact with gauge fields and lead to variations of the fine-structure constant $\alpha$. Here we take advantage of recent observational progress and use a combination of background cosmological observations of Type Ia supernovas and astrophysical and local measurements of $\alpha$ to improve constraints on this class of models. We show that the constraints on $\alpha$ imply that the field dynamics must be extremely slow, leading to a constraint of the present-day dark energy equation of state $(1+w_0)<2.4\times10^{-7}$ at the $99.7\%$ confidence level. Therefore current and forthcoming standard background cosmology observational probes can't distinguish this class of models from a cosmological constant, while detections of $\alpha$ variations could possibly do so since they would have a characteristic redshift dependence.
\end{abstract}
\maketitle

\section{Introduction}

Identifying the physical mechanism behind the recent acceleration of the universe \cite{SN1,SN2} is arguably the most pressing problem of modern physics and cosmology. While current data is broadly in agreement with a cosmological constant, which is also the simplest available explanation (at least in the sense of requiring the smallest number of additional parameters), such an explanation comes with the cost of significant and well-known fine-tuning problems. It is therefore essential to explore possible alternative theoretical scenarios while simultaneously identifying new observational probes that can lead to a more detailed characterization of the properties of the dark side of the universe and to discriminating tests between competing paradigms.

The most obvious alternative to a cosmological constant consists of invoking dynamical degrees of freedom, of which scalar fields are the simplest realization. If such fields are indeed present, one expects them to couple to the rest of the model, unless a yet-unknown symmetry is postulated to suppress these couplings \cite{Carroll,Dvali,Chiba}. In particular, a coupling of the field to the electromagnetic sector will lead to spacetime variations of the fine-structure constant $\alpha$---see \cite{uzanLR,cjmGRG} for recent reviews on this topic. There are some indications of such a variation \cite{Webb}, at the relative level of variation of a few parts per million and in the approximate redshift range $1<z<3$. An ongoing dedicated Large Program at ESO's Very Large Telescope (VLT) is aiming to test them \cite{LP1,LP3}, and the next generation of high-resolution ultra-stable spectrographs will significantly improve the sensitivity of these tests.

Regardless of the outcome of these studies (i.e., whether they provide detections of variations or just null results) these measurements have cosmological implications that go beyond the mere fundamental nature of the tests themselves. These have been recently explored using currently available data \cite{Pinho,Pinho2,Pinho3}, and forecasts for various future facility scenarios have been discussed in some detail \cite{Amendola,LeiteNEW}. These previous studies mostly focused on canonical scalar fields. However, the techniques developed therein are more generic, and here we will exploit them in the context of a different class of models.

Constraints on Dirac-Born-Infeld (DBI) type dark energy models from varying $\alpha$ have first been discussed in \cite{Garousi}. They point out that the DBI action based on string theory naturally gives rise to a coupling between gauge fields and a scalar field responsible for the universe's acceleration. In other words, the field dynamics itself leads to $\alpha$ variations. They place constraints on specific choices of potentials, finding that some fine-tuning is needed for the potentials they consider: the potentials must be quite flat. Here we extend this analysis by exploiting the availability of additional data, but also carry out the analysis for more generic potentials and provide additional insight into the physical interpretation and relevance of the resulting constraints.

A rolling tachyon is an example of a Born-Infeld scalar, and these are well motivated in string theory \cite{Sen1,Sen2}. The interaction of scalar fields with gauge fields will naturally lead to fine-structure constant variations. A further relevant difference is that whereas the coupling of a quintessence field to matter and radiation is not fixed by the standard model of particle physics, these models provide an example where the form of these couplings can be obtained more directly from a fundamental theory, specifically from an effective D-brane action \cite{Garousi}. Therefore, apart form their intrinsic interest, they are also useful as a benchmark to study the discriminating power of future facilities among different classes of models since, as we will show, they do have some interesting distinguishing features.

\section{Cosmological and astrophysical datasets}
\label{sec:data}

We will constrain tachyon dark energy models by using a slightly extended version of the datasets that were also used in \cite{Pinho,Pinho2,Pinho3}, as follows
\begin{itemize}
\item Cosmological data: we use the Union2.1 dataset of 580 Type Ia supernovas \cite{Union} as well as a set of 35 Hubble parameter measurement, of which 28 are described in the compilation of Farooq \& Ratra \cite{Farooq} while 7 more recent ones come from the work of Moresco {\it et al.} \cite{Moresco1,Moresco2}. We will assume that the observations leading to these datasets are not affected by possible $\alpha$ variations. While a varying $\alpha$ is known to affect the luminosity of Type Ia supernovas, a recent analysis shows  \cite{Erminia2} that for parts-per-million level $\alpha$ variations the effect is too small to have an impact on current datasets. As will be shown in what follows, this data will mainly constrain the matter density of the universe, effectively providing us with a prior on it.
\item Laboratory data: we will use the atomic clock constraint on the current drift of $\alpha$ of Rosenband {\it et al.} \cite {Rosenband},
\begin{equation} \label{clocks0}
\frac{\dot\alpha}{\alpha} =(-1.6\pm2.3)\times10^{-17}\,{\rm yr}^{-1}\,.
\end{equation}
which we can also write in a dimensionless form by dividing by the present-day Hubble parameter,
\begin{equation} \label{clocks}
\frac{1}{H_0}\frac{\dot\alpha}{\alpha} =(-2.2\pm3.2)\times10^{-7}\,.
\end{equation}
This is the strongest available laboratory constraint on $\alpha$ only. Other existing laboratory constraints are weaker and also depend on other couplings. (The interested reader can find a summery of other atomic clock tests in \cite{Ferreira2}.) Additionally we will consider the constraint from the Oklo natural nuclear reactor \cite{Oklo}
\begin{equation} \label{okloalpha}
\frac{\Delta\alpha}{\alpha} =(0.5\pm6.1)\times10^{-8}\,,
\end{equation}
at an effective redshift $z=0.14$, though it turns out that for this class of models the atomic clock measurement is more constraining.
\item Astrophysical data: we will use both the spectroscopic measurements of $\alpha$ of Webb {\it et al.} \cite{Webb} (a large dataset of 293 archival data measurements) and the smaller but more recent dataset of 11 dedicated measurements listed in Table \ref{table1}. The latter include the early results of the UVES Large Program for Testing Fundamental Physics \cite{LP1,LP3}, which is expected to be the one with a better control of possible systematics.
\end{itemize}

\begin{table}
\centering
\begin{tabular}{|c|c|c|c|c|}
\hline
 Object & z & ${ \Delta\alpha}/{\alpha}$ (ppm) & Spectrograph & Ref. \\ 
\hline\hline
3 sources & 1.08 & $4.3\pm3.4$ & HIRES & \protect\cite{Songaila} \\
\hline
HS1549$+$1919 & 1.14 & $-7.5\pm5.5$ & UVES/HIRES/HDS & \protect\cite{LP3} \\
\hline
HE0515$-$4414 & 1.15 & $-0.1\pm1.8$ & UVES & \protect\cite{alphaMolaro} \\
\hline
HE0515$-$4414 & 1.15 & $0.5\pm2.4$ & HARPS/UVES & \protect\cite{alphaChand} \\
\hline
HS1549$+$1919 & 1.34 & $-0.7\pm6.6$ & UVES/HIRES/HDS & \protect\cite{LP3} \\
\hline
HE0001$-$2340 & 1.58 & $-1.5\pm2.6$ &  UVES & \protect\cite{alphaAgafonova}\\
\hline
HE1104$-$1805A & 1.66 & $-4.7\pm5.3$ & HIRES & \protect\cite{Songaila} \\
\hline
HE2217$-$2818 & 1.69 & $1.3\pm2.6$ &  UVES & \protect\cite{LP1}\\
\hline
HS1946$+$7658 & 1.74 & $-7.9\pm6.2$ & HIRES & \protect\cite{Songaila} \\
\hline
HS1549$+$1919 & 1.80 & $-6.4\pm7.2$ & UVES/HIRES/HDS & \protect\cite{LP3} \\
\hline
Q1101$-$264 & 1.84 & $5.7\pm2.7$ &  UVES & \protect\cite{alphaMolaro}\\
\hline
\end{tabular}
\caption{\label{table1}Recent dedicated measurements of $\alpha$. Listed are, respectively, the object along each line of sight, the redshift of the measurement, the measurement itself (in parts per million), the spectrograph, and the original reference. The first measurement is the weighted average from 8 absorbers in the redshift range $0.73<z<1.53$ along the lines of sight of HE1104-1805A, HS1700+6416 and HS1946+7658, reported in \cite{Songaila} without the values for individual systems. The UVES, HARPS, HIRES and HDS spectrographs are respectively in the VLT, ESO 3.6m, Keck and Subaru telescopes.}
\end{table}

Our main interest in the present work is to constrain the coupling of the field to the electromagnetic sector. As we will see in the next section, in this class of models this is equivalent to constraining the dark energy equation of state---and consequently also the shape of the potential. For this reason we will fix the Hubble parameter to be $H_0=70$ km/s/Mpc, while this coupling and the matter density will be our free parameters. For simplicity we further assume a flat universe, so $\Omega_m+\Omega_{\phi}=1$, and the $\Omega_i$ denote the present day values. These choices are fully consistent with the cosmological datasets we use, and also with constraints from the cosmic microwave background \cite{Planck}.

\section{Tachyon dark energy models}

The tree-level D-brane action is a Dirac-Born-Infeld type action containing both gauge fields and scalar fields such as tachyons \cite{Sen1,Sen2}, and this action naturally gives rise to the coupling of the Born-Infeld scalars to the gauge fields, which can account for a varying $\alpha$. Rolling tachyon fields naturally arise in string theory, as discussed in \cite{Sen1,Sen2}, and they have been suggested as a candidate to explain the acceleration of the universe \cite{Sen1}. The cosmology of a homogeneous tachyon scalar field as dark energy was first studied in \cite{Bagla}, and the $\alpha$ variation for a Born-Infeld scalar coupled to the gauge field has been previously discussed in  \cite{Garousi}, who obtain some qualitative constraints which will be further quantified by us. Here we first summarize and then extend both of these analyses.

\subsection{A slow-roll tachyon parametrization}

We start by focusing on the tachyon part of the DBI action. Generically its Lagrangian can be written
\be
{\cal L_{\rm tac}}=-V(\phi)\sqrt{1-\partial_a\phi\partial^a\phi}\,,
\ee
with the energy density and pressure being given by
\be
\rho_\phi=\frac{V(\phi)}{\sqrt{1-\partial_a\phi\partial^a\phi}}
\ee
\be
p_\phi=-V(\phi)\sqrt{1-\partial_a\phi\partial^a\phi}\,.
\ee
We will consider the case of a homogeneous field in a Friedmann-Lemaitre-Robertson-Walker background, containing also matter. In that case we have
\be
H^2=\frac{8\pi G}{3}(\rho_m+\rho_\phi)
\ee
and
\be
\frac{\ddot \phi}{1-{\dot\phi}^2}+3H{\dot\phi}+\frac{1}{V}\frac{dV}{d\phi}=0\,.
\ee
Note that the tachyon field equation of state and sound speed are
\be
w_\phi={\dot\phi}^2-1\ge-1\,,
\ee
\be
c^2_s=1-{\dot\phi}^2\le1\,;
\ee
it is also useful to write
\be
{\dot\rho_\phi}=-3H(1+w_\phi)\rho_\phi=-3H\rho_\phi {\dot\phi}^2\,.
\ee
In the case where the tachyon is the single component (ie, neglecting matter as well as radiation) there is a well-known solution \cite{Thanu}
\be
a\propto t^n
\ee
\be
\phi=\sqrt{\frac{2}{3n}}\, t
\ee
which ensues for the potential
\be
V(\phi)=\frac{n}{4\pi G}\left(1-\frac{2}{3n}\right)^{1/2}\frac{1}{\phi^2}\,.
\ee

Now, we start by noting that in these models the field is constrained to be slow-rolling (especially so if it induces $\alpha$ variations, as we will shortly confirm), and in that case the scalar field equation can be approximated to
\be
3H{\dot\phi}\propto -\frac{d\ln{V}}{d\phi}\,.
\ee
Moreover, the right-hand side of this equation is a function of the field $\phi$ and the field is approximately constant. We can thus Taylor-expand the field, and write the Friedmann equation as follows
\be
\frac{H^2}{H_0^2}=\Omega_m(1+z)^3+\Omega_\phi\left[1+\left(\frac{V'}{V}\right)_0(\phi-\phi_0)\right]
\ee
with, from the scalar field equation,
\be
(\phi-\phi_0)= -\frac{1}{3}\left(\frac{1}{H}\frac{V'}{V}\right)_0(t-t_0)\,.
\ee
We therefore have
\be
\frac{H^2}{H_0^2}=\Omega_m(1+z)^3+(1-\Omega_m)\left[1-\frac{1}{3H_0}\left(\frac{V'}{V}\right)^2_0(t-t_0)\right]\,,
\ee
where we also used $\Omega_m+\Omega_\phi=1$. Now, given the slow-roll approximation the correction term in square brackets is expected to be small, and therefore the calculation of the $(t-t_0)$ term can be done assuming the $\Lambda$CDM limit (in other words, the differences will be of higher order), which allows an analytic calculation to be done. After some algebra we find
\be
\frac{H^2}{H_0^2}=\Omega_m(1+z)^3+(1-\Omega_m)\left[1+\frac{2}{9}\lambda^2 f(\Omega_m,z)\right],,
\ee
where we have defined the dynamically relevant dimensionless parameter
\be
\lambda=\frac{1}{H_0}\left(\frac{V'}{V}\right)_0
\ee
and the redshift-dependent correction factor is
\begin{widetext}
\be
f(\Omega_m,z)=\frac{1}{\sqrt{1-\Omega_m}}\ln{\frac{(1+\sqrt{1-\Omega_m})(1+z)^{3/2}}{\sqrt{1-\Omega_m}+\sqrt{\Omega_m(1+z)^3+1-\Omega_m}}}\,.
\ee

It is also useful to calculate the dark energy equation of state in these models. This can be straightforwardly done using the relation
\be
\frac{d\rho_\phi}{dz}=3\frac{1+w_\phi}{1+z}\rho_\phi
\ee
and leads to
\be
1+w_\phi={\dot\phi^2}=\frac{\lambda^2}{9+2\lambda^2f(\Omega_m,z)}\frac{\sqrt{1-\Omega_m}+\sqrt{E(\Omega_m,z)}}{E(\Omega_m,z)+\sqrt{(1-\Omega_m)E(\Omega_m,z)}},,
\ee
\end{widetext}
where for convenience we also defined
\be
E(\Omega_m,z)=\Omega_m(1+z)^3+1-\Omega_m\,.
\ee
As expected the field speed parametrizes the deviation of the dark energy equation of state from the cosmological constant value. Note that this equation of state $(1+w_\phi)$ tends to zero at high redshifts; in other words, these are thawing dark energy models. In particular, the equation of state at the present day is
\be
1+w_0={\dot\phi^2}_0=\frac{\lambda^2}{9}\,,
\ee
providing further physical insight into the role of the parameter $\lambda$.

\subsection{Time variation of $\alpha$}

We now turn to the interaction part of the DBI Lagrangian which is responsible for the $\alpha$ variation. This has the form \cite{Sen1,Sen2,Garousi}
\be
{\cal L_{\rm int}}=\frac{(2\pi{\alpha_s}')^2}{4\beta^2}V(\phi)Tr(g^{-1}Fg^{-1}F)+\ldots \,,
\ee
where $g$ and $F$ are the traces of the four-dimensional metric and the Maxwell tensor respectively, ${\alpha_s}'$ (not to be confused with the fine-structure constant) is related to the string mass scale via $M_s=1/\sqrt{{\alpha_s}'}$, and $\beta$ is a warped factor. (We note that the DBI Lagrangian contains further terms that are of similar order in the gauge field, but these are not relevant for our work since they do not contribute to the $\alpha$ variation. A more systematic discussion can be found in \cite{Sen1,Sen2}.)

This implies, by comparison to the standard Yang-Mills case, that the value of the fine-structure constant in this case is
\be
\alpha(\phi)=\frac{\beta^2M_s^4}{2\pi}\frac{1}{V(\phi)}\,,
\ee
and therefore in these models the fine-structure constant is inversely proportional to the tachyon potential. Expressing this in terms of the relative variation of $\alpha$ with respect to the present day, we finally obtain
\be
\frac{\Delta\alpha}{\alpha}(z)\equiv\frac{\alpha(z)-\alpha_0}{\alpha_0}=\frac{V(\phi_0)}{V(\phi)}-1\,,
\ee
with $\alpha_0\sim1/137$ being the present-day value. Thus a negative value of $\Delta\alpha/\alpha$ corresponds to a smaller value of $\alpha$ in the past (meaning a weaker electromagnetic interaction), which in this class of models corresponds to a larger value of the potential $V(\phi)$.

Given this explicit dependence on the scalar field potential we can now use the same Taylor expansion of the previous subsection, and re-write this as
\be
\frac{\Delta\alpha}{\alpha}\simeq -\left(\frac{V'}{V}\right)_0(\phi-\phi_0)\simeq \frac{1}{3H_0}\left(\frac{V'}{V}\right)^2_0(t-t_0)\,.
\ee
This implies that in these models the fine-structure constant is always smaller in the past (and varies approximately linearly in time). Finally we can write
\be
\frac{\Delta\alpha}{\alpha}=-\frac{2}{9}\lambda^2f(\Omega_m,z)\,,
\ee
which shows that the dimensionless parameter $\lambda$ also provides the overall normalization for this variation. We could even write the suggestive
\be
\frac{H^2}{H_0^2}=\Omega_m(1+z)^3+(1-\Omega_m)\left[1-\frac{\Delta\alpha}{\alpha}(z)\right]\,.
\ee
This makes it clear that in this class of models any deviations from the $\Lambda$CDM behavior must be small, as we now further quantify.

Indeed, we can trivially write the present-day rate of change of the fine-structure constant
\be
\frac{1}{H_0}\left(\frac{\dot\alpha}{\alpha}\right)_0= \frac{1}{3H_0^2}\left(\frac{V'}{V}\right)^2_0\,,
\ee
or equivalently, in terms of the present day dark energy equation of state
\be
\frac{1}{H_0}\left(\frac{\dot\alpha}{\alpha}\right)_0=\frac{1}{3}\lambda^2=3{\dot\phi_0}^2=3(1+w_0)\,.
\ee
Now, as pointed out in the previous section this drift rate is constrained by laboratory measurements with atomic clocks \cite{Rosenband}
\be
\frac{1}{H_0}\left(\frac{\dot\alpha}{\alpha}\right)_0=(-2.2\pm3.2)\times10^{-7}\,,
\ee
showing that in these models $w_0$ is effectively indistinguishable from a cosmological constant, although they can have a distinctive astrophysical variation of $\alpha$. In this sense these models are effectively a physical realization of the more phenomenological Bekenstein-Sandvik-Barrow-Magueijo class of models \cite{BSBM}. This constraint also implies that the field speed today must be tiny
\be
{\dot\phi_0}\leq 10^{-3}\,,
\ee
justifying our slow-roll approximation and also motivating the choice of a logarithmic prior for $\lambda$.

Despite the fact that our analysis is generic (with the relevant information on the shape of the potential being encapsulated in the parameter $\lambda$), we can as an exercise compute $\lambda$ and the $\alpha$ variation for the three specific classes of models that were considered in \cite{Garousi}. For the exponential potential
\be
V(\phi)=V_0 e^{-\mu\phi}
\ee
we have
\be
\lambda=-\frac{\mu}{H_0}
\ee
and
\be
\frac{\Delta\alpha}{\alpha}=e^{\mu(\phi-\phi_0)}\simeq\mu(\phi-\phi_0)
\ee
while for the inverse polynomial potential
\be
V(\phi)=M^{4-n}\phi^{-n}
\ee
we have
\be
\lambda=-\frac{n}{H_0\phi_0}
\ee
and
\be
\frac{\Delta\alpha}{\alpha}=\left(\frac{\phi}{\phi_0}\right)^n-1\simeq\frac{n}{\phi_0}(\phi-\phi_0)
\ee
and finally for the massive rolling scalar potential
\be
V(\phi)=V_0 e^{\frac{1}{2}M^2\phi^2}
\ee
we have
\be
\lambda=\frac{M^2\phi_0}{H_0}
\ee
and
\be
\frac{\Delta\alpha}{\alpha}=e^{\frac{M^2}{2}(\phi_0^2-\phi^2)}\simeq-\frac{M^2}{2}(\phi^2-\phi_0^2)\simeq-M^2\phi_0(\phi-\phi_0)
\ee
as expected.

\section{Constraints on the model}

The work of \cite{Bagla} does a simple comparison with early Type Ia supernova observations. Here we will extend this, using both the more recent Union2.1 supernova dataset and also a set of Hubble parameter measurements discussed in the previous sections \cite{Farooq,Moresco1,Moresco2}. We assume a flat universe and carry out a two-parameter analysis $(\Omega_m,\lambda)$ with a flat prior on the former and a logarithmic prior on the latter. In principle we could include $H_0$ as a third parameter, but we note that the Union2.1 dataset we use already has $H_0=70$ km/s/Mpc. For the fine-structure constant measurements we will use the aforementioned datasets.

\begin{figure*}[!]
\centering
\includegraphics[width=3in]{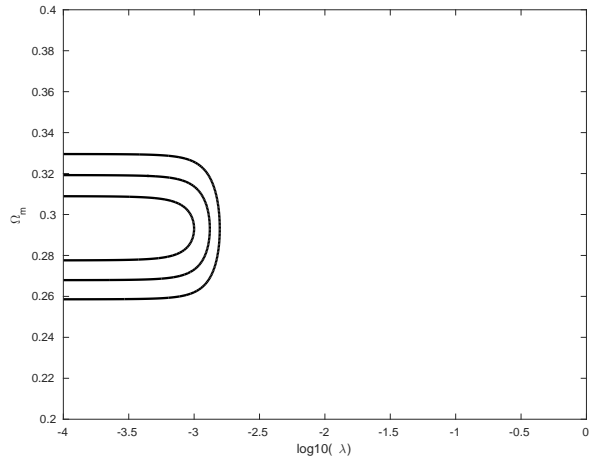}
\caption{\label{fig2Dlike}2D likelihood in the $\lambda$-$\Omega_m$ plane, for the combination of the cosmological, astrophysical and laboratory datasets. One, two and three sigma contours are shown.}
\end{figure*}
\begin{figure}[!]
\centering
\includegraphics[width=3in]{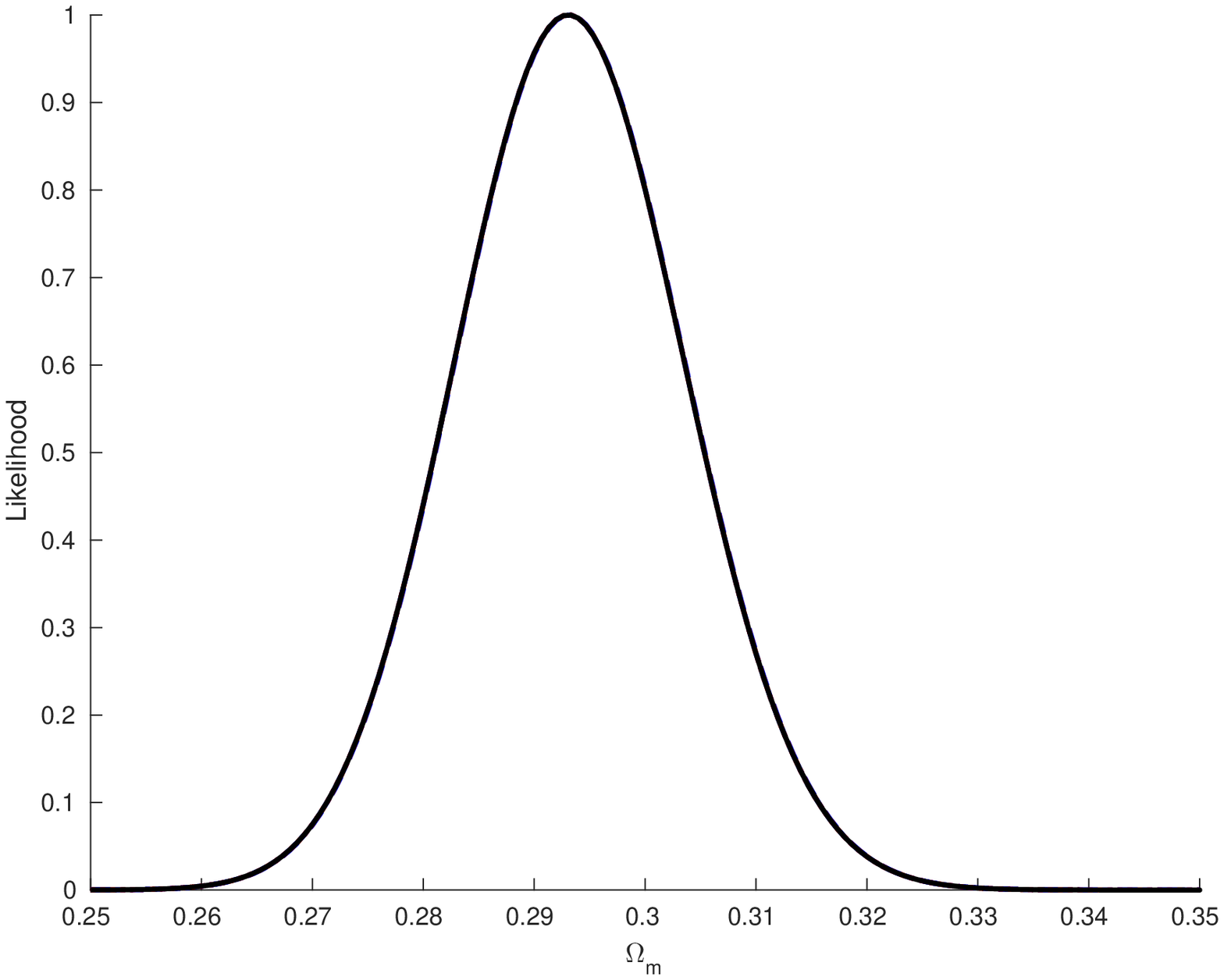}
\includegraphics[width=3in]{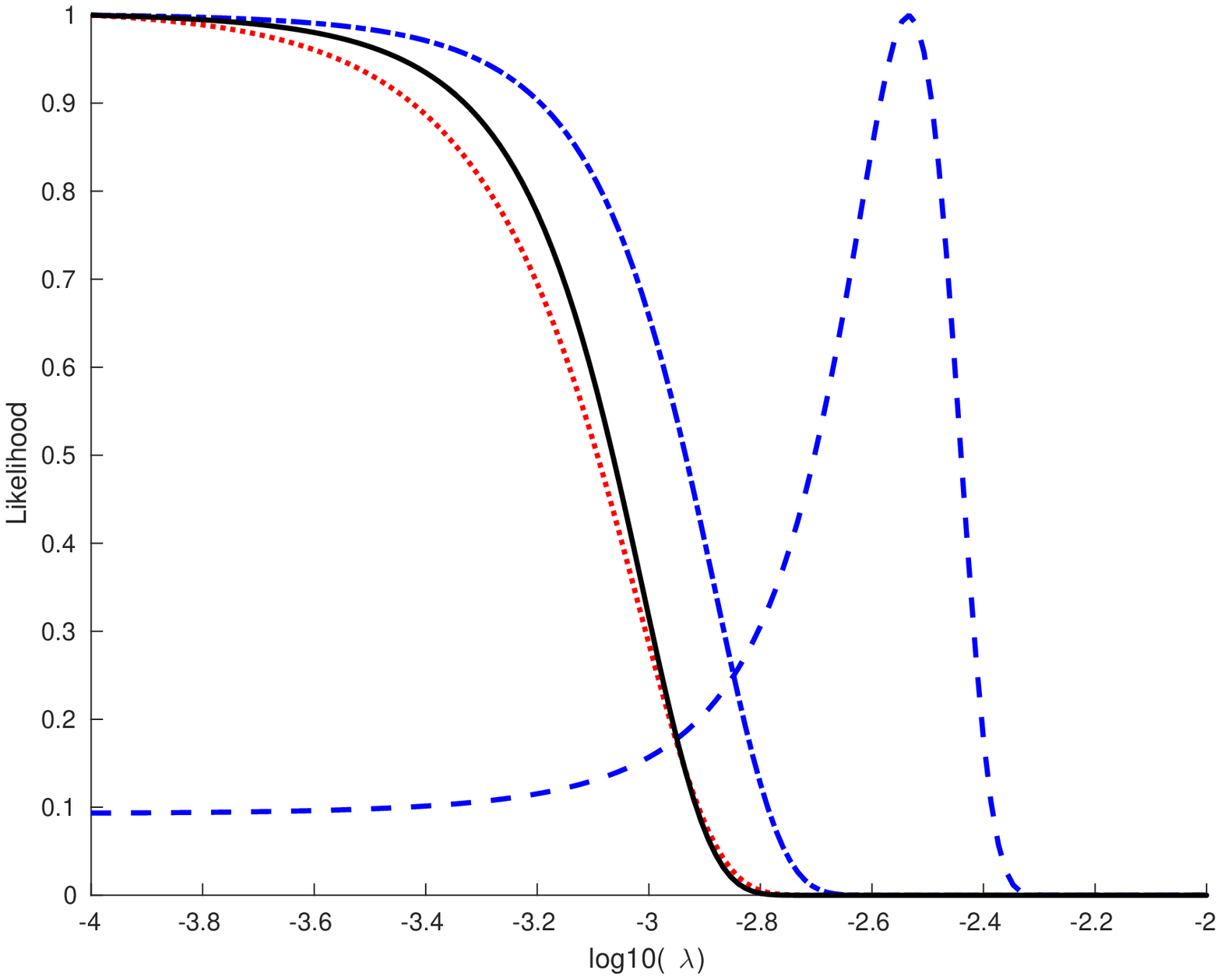}
\caption{\label{fig1Like}1D marginalized likelihood for $\Omega_m$ (top panel) and $\lambda$ (bottom panel. In both cases the blue dashed lines correspond to the combination of cosmological and Webb {\it et al.} data, the blue dash-dotted line corresponds to the combination of cosmological, Table I and Oklo data, the red dotted line corresponds to the combination of cosmological and atomic clock data, and the black solid line corresponds to the combination of all datasets.}
\end{figure}

Our results are summarized in Figs. \ref{fig2Dlike} and \ref{fig1Like}. As expected the cosmological datasets fix the matter density, with the $\alpha$ measurements having very little impact on it since the dependence is only logarithmic. Specifically, marginalizing over $\lambda$ we find the following constraint
\be
\Omega_m==0.29\pm0.03\,,
\ee
at the three sigma ($99.7\%$) confidence level, which is fully compatible with other extant cosmological datasets. On the other hand, the $\alpha$ measurements strongly constrain $\lambda$, for the reasons already explained in the previous section. In particular we notice that the Webb {\it et al.} dataset leads to a two-sigma detection of a non-zero $\lambda$, but the coupling is consistent with zero for other measurements of $\alpha$ and also for the combination of all the datasets we studied. In this case we find, marginalizing over $\Omega_m$,
\be
\lambda<7.8\times10^{-4}\,,\quad 68.3\% C.L.
\ee
\be
\lambda<1.5\times10^{-3}\,,\quad 99.7\% C.L.\,.
\ee
In particular, this leads to an extremely strong constraint on the value of the present day dark energy equation of state
\be
(1+w_0)<2.4\times10^{-7}\,,\quad 99.7\% C.L.\,.
\ee

It is clear that neither current nor foreseen standard probes of background cosmology will be unable to detect such a small deviation from $w_0=-1$. Thus the only possibilities to distinguish these models from the $\Lambda$CDM paradigm would be to rely on their clustering properties (whose study is left for subsequent work) or to use astrophysical measurements of the redshift dependence of $\alpha$.

\section{Conclusions}

We used a combination of astrophysical spectroscopy and local laboratory tests of the stability of the fine-structure constant $\alpha$, complemented by background cosmological datesets, to constrain a class of rolling tachyon models. Part of the motivation for these models stems from the fact, emphasized for example in \cite{Bagla}, that the tachyon Lagrangian generalizes the one for a relativistic particle, just like the one for quintessence generalizes that for a non-relativistic particle. Moreover they are well motivated from string theory, and they naturally couple to gauge fields in a calculable way, in particular leading to a variation of the fine-structure constant $\alpha$.

At the phenomenological level the interesting feature of these models is that a single parameter---effectively the steepness of the potential, in dimensionless units---determines both the dark energy equation of state and the overall level of the $\alpha$ variations. Moreover, these are necessarily thawing models with a monotonically increasing value of $\alpha$ (in other words, smaller values of $\alpha$ in the past). The current local and astrophysical tests of the stability of $\alpha$ therefore place strong constraints on the steepness of the potential, and imply that the present-day value of the dark energy equation of state, although not exactly $-1$, is effectively indistinguishable from it if one restricts oneself to standard observational probes.

Presently these constraints are dominated by the atomic clock tests \cite{Rosenband}, but forthcoming improvements in astrophysical measurements will allow significantly stronger constraints. Specifically the ESPRESSO spectrograph, due for commissioning in the Spring of 2017, and ELT-HIRES, foreseen for the European Extremely Large Telescope, will be ideal for this task. A roadmap for these studies is outlined in \cite{cjmGRG}, and more detailed forecasts of the future impact of these measurements may be found in \cite{LeiteNEW}.

Last but not least, our work demonstrates the importance of testing the stability of nature's fundamental couplings over a broad range of redshifts and accurately mapping their behavior. As this class of rolling tachyon models shows, this may turn out to be the best way we have of identifying deviations from the $\Lambda$CDM paradigm, at least in the next decades. Moreover, in the event of confirmed detections of variations such a mapping is a powerful discriminator, since different classes of models lead to significantly different behaviors for the redshift dependence of $\alpha$---e.g., compare the present models with the canonical ones studied in \cite{Pinho,Pinho2,Pinho3}. We leave a more detailed description of this model selection process for subsequent work.

\begin{acknowledgments}

We are grateful to Ana Catarina Leite and Ana Marta Pinho for helpful discussions on the subject of this work. This work was done in the context of project PTDC/FIS/111725/2009 (FCT, Portugal), with additional support from grant UID/FIS/04434/2013. CJM is also supported by an FCT Research Professorship, contract reference IF/00064/2012, funded by FCT/MCTES (Portugal) and POPH/FSE (EC). FM is supported by grant 201660/2014-8 from CNPq (Brazil).

CJM thanks the Galileo Galilei Institute for Theoretical Physics for the hospitality and the INFN for partial support during the completion of this work.

\end{acknowledgments}

\bibliography{tachyon}

\begin{thebibliography}{34}%
\makeatletter
\providecommand \@ifxundefined [1]{%
 \@ifx{#1\undefined}
}%
\providecommand \@ifnum [1]{%
 \ifnum #1\expandafter \@firstoftwo
 \else \expandafter \@secondoftwo
 \fi
}%
\providecommand \@ifx [1]{%
 \ifx #1\expandafter \@firstoftwo
 \else \expandafter \@secondoftwo
 \fi
}%
\providecommand \natexlab [1]{#1}%
\providecommand \enquote  [1]{``#1''}%
\providecommand \bibnamefont  [1]{#1}%
\providecommand \bibfnamefont [1]{#1}%
\providecommand \citenamefont [1]{#1}%
\providecommand \href@noop [0]{\@secondoftwo}%
\providecommand \href [0]{\begingroup \@sanitize@url \@href}%
\providecommand \@href[1]{\@@startlink{#1}\@@href}%
\providecommand \@@href[1]{\endgroup#1\@@endlink}%
\providecommand \@sanitize@url [0]{\catcode `\\12\catcode `\$12\catcode
  `\&12\catcode `\#12\catcode `\^12\catcode `\_12\catcode `\%12\relax}%
\providecommand \@@startlink[1]{}%
\providecommand \@@endlink[0]{}%
\providecommand \url  [0]{\begingroup\@sanitize@url \@url }%
\providecommand \@url [1]{\endgroup\@href {#1}{\urlprefix }}%
\providecommand \urlprefix  [0]{URL }%
\providecommand \Eprint [0]{\href }%
\providecommand \doibase [0]{http://dx.doi.org/}%
\providecommand \selectlanguage [0]{\@gobble}%
\providecommand \bibinfo  [0]{\@secondoftwo}%
\providecommand \bibfield  [0]{\@secondoftwo}%
\providecommand \translation [1]{[#1]}%
\providecommand \BibitemOpen [0]{}%
\providecommand \bibitemStop [0]{}%
\providecommand \bibitemNoStop [0]{.\EOS\space}%
\providecommand \EOS [0]{\spacefactor3000\relax}%
\providecommand \BibitemShut  [1]{\csname bibitem#1\endcsname}%
\let\auto@bib@innerbib\@empty
\bibitem [{\citenamefont {Riess}\ \emph {et~al.}(1998)\citenamefont {Riess}
  \emph {et~al.}}]{SN1}%
  \BibitemOpen
  \bibfield  {author} {\bibinfo {author} {\bibfnamefont {A.~G.}\ \bibnamefont
  {Riess}} \emph {et~al.} (\bibinfo {collaboration} {Supernova Search Team}),\
  }\href {\doibase 10.1086/300499} {\bibfield  {journal} {\bibinfo  {journal}
  {Astron.J.}\ }\textbf {\bibinfo {volume} {116}},\ \bibinfo {pages} {1009}
  (\bibinfo {year} {1998})},\ \Eprint {http://arxiv.org/abs/astro-ph/9805201}
  {arXiv:astro-ph/9805201 [astro-ph]} \BibitemShut {NoStop}%
\bibitem [{\citenamefont {Perlmutter}\ \emph {et~al.}(1999)\citenamefont
  {Perlmutter} \emph {et~al.}}]{SN2}%
  \BibitemOpen
  \bibfield  {author} {\bibinfo {author} {\bibfnamefont {S.}~\bibnamefont
  {Perlmutter}} \emph {et~al.} (\bibinfo {collaboration} {Supernova Cosmology
  Project}),\ }\href {\doibase 10.1086/307221} {\bibfield  {journal} {\bibinfo
  {journal} {Astrophys.J.}\ }\textbf {\bibinfo {volume} {517}},\ \bibinfo
  {pages} {565} (\bibinfo {year} {1999})},\ \Eprint
  {http://arxiv.org/abs/astro-ph/9812133} {arXiv:astro-ph/9812133 [astro-ph]}
  \BibitemShut {NoStop}%
\bibitem [{\citenamefont {Carroll}(1998)}]{Carroll}%
  \BibitemOpen
  \bibfield  {author} {\bibinfo {author} {\bibfnamefont {S.~M.}\ \bibnamefont
  {Carroll}},\ }\href {\doibase 10.1103/PhysRevLett.81.3067} {\bibfield
  {journal} {\bibinfo  {journal} {Phys.Rev.Lett.}\ }\textbf {\bibinfo {volume}
  {81}},\ \bibinfo {pages} {3067} (\bibinfo {year} {1998})},\ \Eprint
  {http://arxiv.org/abs/astro-ph/9806099} {arXiv:astro-ph/9806099 [astro-ph]}
  \BibitemShut {NoStop}%
\bibitem [{\citenamefont {Dvali}\ and\ \citenamefont
  {Zaldarriaga}(2002)}]{Dvali}%
  \BibitemOpen
  \bibfield  {author} {\bibinfo {author} {\bibfnamefont {G.}~\bibnamefont
  {Dvali}}\ and\ \bibinfo {author} {\bibfnamefont {M.}~\bibnamefont
  {Zaldarriaga}},\ }\href {\doibase 10.1103/PhysRevLett.88.091303} {\bibfield
  {journal} {\bibinfo  {journal} {Phys.Rev.Lett.}\ }\textbf {\bibinfo {volume}
  {88}},\ \bibinfo {pages} {091303} (\bibinfo {year} {2002})},\ \Eprint
  {http://arxiv.org/abs/hep-ph/0108217} {arXiv:hep-ph/0108217 [hep-ph]}
  \BibitemShut {NoStop}%
\bibitem [{\citenamefont {Chiba}\ and\ \citenamefont {Kohri}(2002)}]{Chiba}%
  \BibitemOpen
  \bibfield  {author} {\bibinfo {author} {\bibfnamefont {T.}~\bibnamefont
  {Chiba}}\ and\ \bibinfo {author} {\bibfnamefont {K.}~\bibnamefont {Kohri}},\
  }\href {\doibase 10.1143/PTP.107.631} {\bibfield  {journal} {\bibinfo
  {journal} {Prog. Theor. Phys.}\ }\textbf {\bibinfo {volume} {107}},\ \bibinfo
  {pages} {631} (\bibinfo {year} {2002})},\ \Eprint
  {http://arxiv.org/abs/hep-ph/0111086} {arXiv:hep-ph/0111086 [hep-ph]}
  \BibitemShut {NoStop}%
\bibitem [{\citenamefont {Uzan}(2011)}]{uzanLR}%
  \BibitemOpen
  \bibfield  {author} {\bibinfo {author} {\bibfnamefont {J.-P.}\ \bibnamefont
  {Uzan}},\ }\href@noop {} {\bibfield  {journal} {\bibinfo  {journal} {Living
  Rev.Rel.}\ }\textbf {\bibinfo {volume} {14}},\ \bibinfo {pages} {2} (\bibinfo
  {year} {2011})},\ \Eprint {http://arxiv.org/abs/1009.5514} {arXiv:1009.5514
  [astro-ph.CO]} \BibitemShut {NoStop}%
\bibitem [{\citenamefont {Martins}(2014)}]{cjmGRG}%
  \BibitemOpen
  \bibfield  {author} {\bibinfo {author} {\bibfnamefont {C.~J. A.~P.}\
  \bibnamefont {Martins}},\ }\href {\doibase 10.1007/s10714-014-1843-7}
  {\bibfield  {journal} {\bibinfo  {journal} {Gen.Rel.Grav.}\ }\textbf
  {\bibinfo {volume} {47}},\ \bibinfo {pages} {1843} (\bibinfo {year}
  {2014})},\ \Eprint {http://arxiv.org/abs/1412.0108} {arXiv:1412.0108
  [astro-ph.CO]} \BibitemShut {NoStop}%
\bibitem [{\citenamefont {Webb}\ \emph {et~al.}(2011)\citenamefont {Webb},
  \citenamefont {King}, \citenamefont {Murphy}, \citenamefont {Flambaum},
  \citenamefont {Carswell} \emph {et~al.}}]{Webb}%
  \BibitemOpen
  \bibfield  {author} {\bibinfo {author} {\bibfnamefont {J.}~\bibnamefont
  {Webb}}, \bibinfo {author} {\bibfnamefont {J.}~\bibnamefont {King}}, \bibinfo
  {author} {\bibfnamefont {M.}~\bibnamefont {Murphy}}, \bibinfo {author}
  {\bibfnamefont {V.}~\bibnamefont {Flambaum}}, \bibinfo {author}
  {\bibfnamefont {R.}~\bibnamefont {Carswell}},  \emph {et~al.},\ }\href
  {\doibase 10.1103/PhysRevLett.107.191101} {\bibfield  {journal} {\bibinfo
  {journal} {Phys.Rev.Lett.}\ }\textbf {\bibinfo {volume} {107}},\ \bibinfo
  {pages} {191101} (\bibinfo {year} {2011})},\ \Eprint
  {http://arxiv.org/abs/1008.3907} {arXiv:1008.3907 [astro-ph.CO]} \BibitemShut
  {NoStop}%
\bibitem [{\citenamefont {Molaro}\ \emph {et~al.}(2013)\citenamefont {Molaro},
  \citenamefont {Centurion}, \citenamefont {Whitmore}, \citenamefont {Evans},
  \citenamefont {Murphy} \emph {et~al.}}]{LP1}%
  \BibitemOpen
  \bibfield  {author} {\bibinfo {author} {\bibfnamefont {P.}~\bibnamefont
  {Molaro}}, \bibinfo {author} {\bibfnamefont {M.}~\bibnamefont {Centurion}},
  \bibinfo {author} {\bibfnamefont {J.}~\bibnamefont {Whitmore}}, \bibinfo
  {author} {\bibfnamefont {T.}~\bibnamefont {Evans}}, \bibinfo {author}
  {\bibfnamefont {M.}~\bibnamefont {Murphy}},  \emph {et~al.},\ }\href
  {\doibase 10.1051/0004-6361/201321351} {\bibfield  {journal} {\bibinfo
  {journal} {Astron.Astrophys.}\ }\textbf {\bibinfo {volume} {555}},\ \bibinfo
  {pages} {A68} (\bibinfo {year} {2013})},\ \Eprint
  {http://arxiv.org/abs/1305.1884} {arXiv:1305.1884 [astro-ph.CO]} \BibitemShut
  {NoStop}%
\bibitem [{\citenamefont {{Evans}}\ \emph {et~al.}(2014)\citenamefont
  {{Evans}}, \citenamefont {{Murphy}}, \citenamefont {{Whitmore}},
  \citenamefont {{Misawa}}, \citenamefont {{Centurion}}, \citenamefont
  {{D'Odorico}}, \citenamefont {{Lopez}}, \citenamefont {{Martins}},
  \citenamefont {{Molaro}}, \citenamefont {{Petitjean}}, \citenamefont
  {{Rahmani}}, \citenamefont {{Srianand}},\ and\ \citenamefont
  {{Wendt}}}]{LP3}%
  \BibitemOpen
  \bibfield  {author} {\bibinfo {author} {\bibfnamefont {T.~M.}\ \bibnamefont
  {{Evans}}}, \bibinfo {author} {\bibfnamefont {M.~T.}\ \bibnamefont
  {{Murphy}}}, \bibinfo {author} {\bibfnamefont {J.~B.}\ \bibnamefont
  {{Whitmore}}}, \bibinfo {author} {\bibfnamefont {T.}~\bibnamefont
  {{Misawa}}}, \bibinfo {author} {\bibfnamefont {M.}~\bibnamefont
  {{Centurion}}}, \bibinfo {author} {\bibfnamefont {S.}~\bibnamefont
  {{D'Odorico}}}, \bibinfo {author} {\bibfnamefont {S.}~\bibnamefont
  {{Lopez}}}, \bibinfo {author} {\bibfnamefont {C.~J.~A.~P.}\ \bibnamefont
  {{Martins}}}, \bibinfo {author} {\bibfnamefont {P.}~\bibnamefont {{Molaro}}},
  \bibinfo {author} {\bibfnamefont {P.}~\bibnamefont {{Petitjean}}}, \bibinfo
  {author} {\bibfnamefont {H.}~\bibnamefont {{Rahmani}}}, \bibinfo {author}
  {\bibfnamefont {R.}~\bibnamefont {{Srianand}}}, \ and\ \bibinfo {author}
  {\bibfnamefont {M.}~\bibnamefont {{Wendt}}},\ }\href {\doibase
  10.1093/mnras/stu1754} {\bibfield  {journal} {\bibinfo  {journal}
  {M.N.R.A.S.}\ }\textbf {\bibinfo {volume} {445}},\ \bibinfo {pages} {128}
  (\bibinfo {year} {2014})}\BibitemShut {NoStop}%
\bibitem [{\citenamefont {Martins}\ and\ \citenamefont {Pinho}(2015)}]{Pinho}%
  \BibitemOpen
  \bibfield  {author} {\bibinfo {author} {\bibfnamefont {C.~J. A.~P.}\
  \bibnamefont {Martins}}\ and\ \bibinfo {author} {\bibfnamefont {A.~M.~M.}\
  \bibnamefont {Pinho}},\ }\href {\doibase 10.1103/PhysRevD.91.103501}
  {\bibfield  {journal} {\bibinfo  {journal} {Phys. Rev.}\ }\textbf {\bibinfo
  {volume} {D91}},\ \bibinfo {pages} {103501} (\bibinfo {year} {2015})},\
  \Eprint {http://arxiv.org/abs/1505.02196} {arXiv:1505.02196 [astro-ph.CO]}
  \BibitemShut {NoStop}%
\bibitem [{\citenamefont {Martins}\ \emph {et~al.}(2015)\citenamefont
  {Martins}, \citenamefont {Pinho}, \citenamefont {Alves}, \citenamefont
  {Pino}, \citenamefont {Rocha},\ and\ \citenamefont {von
  Wietersheim}}]{Pinho2}%
  \BibitemOpen
  \bibfield  {author} {\bibinfo {author} {\bibfnamefont {C.~J. A.~P.}\
  \bibnamefont {Martins}}, \bibinfo {author} {\bibfnamefont {A.~M.~M.}\
  \bibnamefont {Pinho}}, \bibinfo {author} {\bibfnamefont {R.~F.~C.}\
  \bibnamefont {Alves}}, \bibinfo {author} {\bibfnamefont {M.}~\bibnamefont
  {Pino}}, \bibinfo {author} {\bibfnamefont {C.~I. S.~A.}\ \bibnamefont
  {Rocha}}, \ and\ \bibinfo {author} {\bibfnamefont {M.}~\bibnamefont {von
  Wietersheim}},\ }\href {\doibase 10.1088/1475-7516/2015/08/047} {\bibfield
  {journal} {\bibinfo  {journal} {JCAP}\ }\textbf {\bibinfo {volume} {1508}},\
  \bibinfo {pages} {047} (\bibinfo {year} {2015})},\ \Eprint
  {http://arxiv.org/abs/1508.06157} {arXiv:1508.06157 [astro-ph.CO]}
  \BibitemShut {NoStop}%
\bibitem [{\citenamefont {Martins}\ \emph {et~al.}(2016)\citenamefont
  {Martins}, \citenamefont {Pinho}, \citenamefont {Carreira}, \citenamefont
  {Gusart}, \citenamefont {L\'opez},\ and\ \citenamefont {Rocha}}]{Pinho3}%
  \BibitemOpen
  \bibfield  {author} {\bibinfo {author} {\bibfnamefont {C.~J. A.~P.}\
  \bibnamefont {Martins}}, \bibinfo {author} {\bibfnamefont {A.~M.~M.}\
  \bibnamefont {Pinho}}, \bibinfo {author} {\bibfnamefont {P.}~\bibnamefont
  {Carreira}}, \bibinfo {author} {\bibfnamefont {A.}~\bibnamefont {Gusart}},
  \bibinfo {author} {\bibfnamefont {J.}~\bibnamefont {L\'opez}}, \ and\
  \bibinfo {author} {\bibfnamefont {C.~I. S.~A.}\ \bibnamefont {Rocha}},\
  }\href {\doibase 10.1103/PhysRevD.93.023506} {\bibfield  {journal} {\bibinfo
  {journal} {Phys. Rev.}\ }\textbf {\bibinfo {volume} {D93}},\ \bibinfo {pages}
  {023506} (\bibinfo {year} {2016})},\ \Eprint
  {http://arxiv.org/abs/1601.02950} {arXiv:1601.02950 [astro-ph.CO]}
  \BibitemShut {NoStop}%
\bibitem [{\citenamefont {Amendola}\ \emph {et~al.}(2012)\citenamefont
  {Amendola}, \citenamefont {Leite}, \citenamefont {Martins}, \citenamefont
  {Nunes}, \citenamefont {Pedrosa} \emph {et~al.}}]{Amendola}%
  \BibitemOpen
  \bibfield  {author} {\bibinfo {author} {\bibfnamefont {L.}~\bibnamefont
  {Amendola}}, \bibinfo {author} {\bibfnamefont {A.~C.~O.}\ \bibnamefont
  {Leite}}, \bibinfo {author} {\bibfnamefont {C.~J. A.~P.}\ \bibnamefont
  {Martins}}, \bibinfo {author} {\bibfnamefont {N.}~\bibnamefont {Nunes}},
  \bibinfo {author} {\bibfnamefont {P.~O.~J.}\ \bibnamefont {Pedrosa}},  \emph
  {et~al.},\ }\href {\doibase 10.1103/PhysRevD.86.063515} {\bibfield  {journal}
  {\bibinfo  {journal} {Phys.Rev.}\ }\textbf {\bibinfo {volume} {D86}},\
  \bibinfo {pages} {063515} (\bibinfo {year} {2012})},\ \Eprint
  {http://arxiv.org/abs/1109.6793} {arXiv:1109.6793 [astro-ph.CO]} \BibitemShut
  {NoStop}%
\bibitem [{\citenamefont {Leite}\ and\ \citenamefont
  {Martins}(2015)}]{LeiteNEW}%
  \BibitemOpen
  \bibfield  {author} {\bibinfo {author} {\bibfnamefont {A.~C.~O.}\
  \bibnamefont {Leite}}\ and\ \bibinfo {author} {\bibfnamefont {C.~J. A.~P.}\
  \bibnamefont {Martins}},\ }\href {\doibase 10.1103/PhysRevD.91.103519}
  {\bibfield  {journal} {\bibinfo  {journal} {Phys. Rev.}\ }\textbf {\bibinfo
  {volume} {D91}},\ \bibinfo {pages} {103519} (\bibinfo {year} {2015})},\
  \Eprint {http://arxiv.org/abs/1505.05529} {arXiv:1505.05529 [astro-ph.CO]}
  \BibitemShut {NoStop}%
\bibitem [{\citenamefont {Garousi}\ \emph {et~al.}(2005)\citenamefont
  {Garousi}, \citenamefont {Sami},\ and\ \citenamefont {Tsujikawa}}]{Garousi}%
  \BibitemOpen
  \bibfield  {author} {\bibinfo {author} {\bibfnamefont {M.~R.}\ \bibnamefont
  {Garousi}}, \bibinfo {author} {\bibfnamefont {M.}~\bibnamefont {Sami}}, \
  and\ \bibinfo {author} {\bibfnamefont {S.}~\bibnamefont {Tsujikawa}},\ }\href
  {\doibase 10.1103/PhysRevD.71.083005} {\bibfield  {journal} {\bibinfo
  {journal} {Phys. Rev.}\ }\textbf {\bibinfo {volume} {D71}},\ \bibinfo {pages}
  {083005} (\bibinfo {year} {2005})},\ \Eprint
  {http://arxiv.org/abs/hep-th/0412002} {arXiv:hep-th/0412002 [hep-th]}
  \BibitemShut {NoStop}%
\bibitem [{\citenamefont {Sen}(2002{\natexlab{a}})}]{Sen1}%
  \BibitemOpen
  \bibfield  {author} {\bibinfo {author} {\bibfnamefont {A.}~\bibnamefont
  {Sen}},\ }\href {\doibase 10.1088/1126-6708/2002/04/048} {\bibfield
  {journal} {\bibinfo  {journal} {JHEP}\ }\textbf {\bibinfo {volume} {04}},\
  \bibinfo {pages} {048} (\bibinfo {year} {2002}{\natexlab{a}})},\ \Eprint
  {http://arxiv.org/abs/hep-th/0203211} {arXiv:hep-th/0203211 [hep-th]}
  \BibitemShut {NoStop}%
\bibitem [{\citenamefont {Sen}(2002{\natexlab{b}})}]{Sen2}%
  \BibitemOpen
  \bibfield  {author} {\bibinfo {author} {\bibfnamefont {A.}~\bibnamefont
  {Sen}},\ }\href {\doibase 10.1088/1126-6708/2002/07/065} {\bibfield
  {journal} {\bibinfo  {journal} {JHEP}\ }\textbf {\bibinfo {volume} {07}},\
  \bibinfo {pages} {065} (\bibinfo {year} {2002}{\natexlab{b}})},\ \Eprint
  {http://arxiv.org/abs/hep-th/0203265} {arXiv:hep-th/0203265 [hep-th]}
  \BibitemShut {NoStop}%
\bibitem [{\citenamefont {Suzuki}\ \emph {et~al.}(2012)\citenamefont {Suzuki},
  \citenamefont {Rubin}, \citenamefont {Lidman}, \citenamefont {Aldering},
  \citenamefont {Amanullah} \emph {et~al.}}]{Union}%
  \BibitemOpen
  \bibfield  {author} {\bibinfo {author} {\bibfnamefont {N.}~\bibnamefont
  {Suzuki}}, \bibinfo {author} {\bibfnamefont {D.}~\bibnamefont {Rubin}},
  \bibinfo {author} {\bibfnamefont {C.}~\bibnamefont {Lidman}}, \bibinfo
  {author} {\bibfnamefont {G.}~\bibnamefont {Aldering}}, \bibinfo {author}
  {\bibfnamefont {R.}~\bibnamefont {Amanullah}},  \emph {et~al.},\ }\href
  {\doibase 10.1088/0004-637X/746/1/85} {\bibfield  {journal} {\bibinfo
  {journal} {Astrophys.J.}\ }\textbf {\bibinfo {volume} {746}},\ \bibinfo
  {pages} {85} (\bibinfo {year} {2012})},\ \Eprint
  {http://arxiv.org/abs/1105.3470} {arXiv:1105.3470 [astro-ph.CO]} \BibitemShut
  {NoStop}%
\bibitem [{\citenamefont {Farooq}\ and\ \citenamefont {Ratra}(2013)}]{Farooq}%
  \BibitemOpen
  \bibfield  {author} {\bibinfo {author} {\bibfnamefont {O.}~\bibnamefont
  {Farooq}}\ and\ \bibinfo {author} {\bibfnamefont {B.}~\bibnamefont {Ratra}},\
  }\href {\doibase 10.1088/2041-8205/766/1/L7} {\bibfield  {journal} {\bibinfo
  {journal} {Astrophys.J.}\ }\textbf {\bibinfo {volume} {766}},\ \bibinfo
  {pages} {L7} (\bibinfo {year} {2013})},\ \Eprint
  {http://arxiv.org/abs/1301.5243} {arXiv:1301.5243 [astro-ph.CO]} \BibitemShut
  {NoStop}%
\bibitem [{\citenamefont {Moresco}(2015)}]{Moresco1}%
  \BibitemOpen
  \bibfield  {author} {\bibinfo {author} {\bibfnamefont {M.}~\bibnamefont
  {Moresco}},\ }\href {\doibase 10.1093/mnrasl/slv037} {\bibfield  {journal}
  {\bibinfo  {journal} {Mon. Not. Roy. Astron. Soc.}\ }\textbf {\bibinfo
  {volume} {450}},\ \bibinfo {pages} {L16} (\bibinfo {year} {2015})},\ \Eprint
  {http://arxiv.org/abs/1503.01116} {arXiv:1503.01116 [astro-ph.CO]}
  \BibitemShut {NoStop}%
\bibitem [{\citenamefont {Moresco}\ \emph {et~al.}(2016)\citenamefont
  {Moresco}, \citenamefont {Pozzetti}, \citenamefont {Cimatti}, \citenamefont
  {Jimenez}, \citenamefont {Maraston}, \citenamefont {Verde}, \citenamefont
  {Thomas}, \citenamefont {Citro}, \citenamefont {Tojeiro},\ and\ \citenamefont
  {Wilkinson}}]{Moresco2}%
  \BibitemOpen
  \bibfield  {author} {\bibinfo {author} {\bibfnamefont {M.}~\bibnamefont
  {Moresco}}, \bibinfo {author} {\bibfnamefont {L.}~\bibnamefont {Pozzetti}},
  \bibinfo {author} {\bibfnamefont {A.}~\bibnamefont {Cimatti}}, \bibinfo
  {author} {\bibfnamefont {R.}~\bibnamefont {Jimenez}}, \bibinfo {author}
  {\bibfnamefont {C.}~\bibnamefont {Maraston}}, \bibinfo {author}
  {\bibfnamefont {L.}~\bibnamefont {Verde}}, \bibinfo {author} {\bibfnamefont
  {D.}~\bibnamefont {Thomas}}, \bibinfo {author} {\bibfnamefont
  {A.}~\bibnamefont {Citro}}, \bibinfo {author} {\bibfnamefont
  {R.}~\bibnamefont {Tojeiro}}, \ and\ \bibinfo {author} {\bibfnamefont
  {D.}~\bibnamefont {Wilkinson}},\ }\href {\doibase
  10.1088/1475-7516/2016/05/014} {\bibfield  {journal} {\bibinfo  {journal}
  {JCAP}\ }\textbf {\bibinfo {volume} {1605}},\ \bibinfo {pages} {014}
  (\bibinfo {year} {2016})},\ \Eprint {http://arxiv.org/abs/1601.01701}
  {arXiv:1601.01701 [astro-ph.CO]} \BibitemShut {NoStop}%
\bibitem [{\citenamefont {Calabrese}\ \emph {et~al.}(2014)\citenamefont
  {Calabrese}, \citenamefont {Martinelli}, \citenamefont {Pandolfi},
  \citenamefont {Cardone}, \citenamefont {Martins}, \citenamefont {Spiro},\
  and\ \citenamefont {Vielzeuf}}]{Erminia2}%
  \BibitemOpen
  \bibfield  {author} {\bibinfo {author} {\bibfnamefont {E.}~\bibnamefont
  {Calabrese}}, \bibinfo {author} {\bibfnamefont {M.}~\bibnamefont
  {Martinelli}}, \bibinfo {author} {\bibfnamefont {S.}~\bibnamefont
  {Pandolfi}}, \bibinfo {author} {\bibfnamefont {V.~F.}\ \bibnamefont
  {Cardone}}, \bibinfo {author} {\bibfnamefont {C.~J. A.~P.}\ \bibnamefont
  {Martins}}, \bibinfo {author} {\bibfnamefont {S.}~\bibnamefont {Spiro}}, \
  and\ \bibinfo {author} {\bibfnamefont {P.~E.}\ \bibnamefont {Vielzeuf}},\
  }\href {\doibase 10.1103/PhysRevD.89.083509} {\bibfield  {journal} {\bibinfo
  {journal} {Phys. Rev.}\ }\textbf {\bibinfo {volume} {D89}},\ \bibinfo {pages}
  {083509} (\bibinfo {year} {2014})},\ \Eprint {http://arxiv.org/abs/1311.5841}
  {arXiv:1311.5841 [astro-ph.CO]} \BibitemShut {NoStop}%
\bibitem [{\citenamefont {Rosenband}\ \emph {et~al.}(2008)\citenamefont
  {Rosenband}, \citenamefont {Hume}, \citenamefont {Schmidt}, \citenamefont
  {Chou}, \citenamefont {Brusch}, \citenamefont {Lorini}, \citenamefont
  {Oskay}, \citenamefont {Drullinger}, \citenamefont {Fortier}, \citenamefont
  {Stalnaker}, \citenamefont {Diddams}, \citenamefont {Swann}, \citenamefont
  {Newbury}, \citenamefont {Itano}, \citenamefont {Wineland},\ and\
  \citenamefont {Bergquist}}]{Rosenband}%
  \BibitemOpen
  \bibfield  {author} {\bibinfo {author} {\bibfnamefont {T.}~\bibnamefont
  {Rosenband}}, \bibinfo {author} {\bibfnamefont {D.}~\bibnamefont {Hume}},
  \bibinfo {author} {\bibfnamefont {P.}~\bibnamefont {Schmidt}}, \bibinfo
  {author} {\bibfnamefont {C.}~\bibnamefont {Chou}}, \bibinfo {author}
  {\bibfnamefont {A.}~\bibnamefont {Brusch}}, \bibinfo {author} {\bibfnamefont
  {L.}~\bibnamefont {Lorini}}, \bibinfo {author} {\bibfnamefont
  {W.}~\bibnamefont {Oskay}}, \bibinfo {author} {\bibfnamefont
  {R.}~\bibnamefont {Drullinger}}, \bibinfo {author} {\bibfnamefont
  {T.}~\bibnamefont {Fortier}}, \bibinfo {author} {\bibfnamefont
  {J.}~\bibnamefont {Stalnaker}}, \bibinfo {author} {\bibfnamefont
  {S.}~\bibnamefont {Diddams}}, \bibinfo {author} {\bibfnamefont
  {W.}~\bibnamefont {Swann}}, \bibinfo {author} {\bibfnamefont
  {N.}~\bibnamefont {Newbury}}, \bibinfo {author} {\bibfnamefont
  {W.}~\bibnamefont {Itano}}, \bibinfo {author} {\bibfnamefont
  {D.}~\bibnamefont {Wineland}}, \ and\ \bibinfo {author} {\bibfnamefont
  {J.}~\bibnamefont {Bergquist}},\ }\href {\doibase 10.1126/science.1154622}
  {\bibfield  {journal} {\bibinfo  {journal} {Science}\ }\textbf {\bibinfo
  {volume} {319}},\ \bibinfo {pages} {1808} (\bibinfo {year}
  {2008})}\BibitemShut {NoStop}%
\bibitem [{\citenamefont {Ferreira}\ and\ \citenamefont
  {Martins}(2015)}]{Ferreira2}%
  \BibitemOpen
  \bibfield  {author} {\bibinfo {author} {\bibfnamefont {M.~C.}\ \bibnamefont
  {Ferreira}}\ and\ \bibinfo {author} {\bibfnamefont {C.~J. A.~P.}\
  \bibnamefont {Martins}},\ }\href {\doibase 10.1103/PhysRevD.91.124032}
  {\bibfield  {journal} {\bibinfo  {journal} {Phys. Rev.}\ }\textbf {\bibinfo
  {volume} {D91}},\ \bibinfo {pages} {124032} (\bibinfo {year} {2015})},\
  \Eprint {http://arxiv.org/abs/1506.03550} {arXiv:1506.03550 [astro-ph.CO]}
  \BibitemShut {NoStop}%
\bibitem [{\citenamefont {Petrov}\ \emph {et~al.}(2006)\citenamefont {Petrov},
  \citenamefont {Nazarov}, \citenamefont {Onegin}, \citenamefont {Petrov},\
  and\ \citenamefont {Sakhnovsky}}]{Oklo}%
  \BibitemOpen
  \bibfield  {author} {\bibinfo {author} {\bibfnamefont {{\relax Yu}.~V.}\
  \bibnamefont {Petrov}}, \bibinfo {author} {\bibfnamefont {A.~I.}\
  \bibnamefont {Nazarov}}, \bibinfo {author} {\bibfnamefont {M.~S.}\
  \bibnamefont {Onegin}}, \bibinfo {author} {\bibfnamefont {V.~{\relax Yu}.}\
  \bibnamefont {Petrov}}, \ and\ \bibinfo {author} {\bibfnamefont {E.~G.}\
  \bibnamefont {Sakhnovsky}},\ }\href {\doibase 10.1103/PhysRevC.74.064610}
  {\bibfield  {journal} {\bibinfo  {journal} {Phys. Rev.}\ }\textbf {\bibinfo
  {volume} {C74}},\ \bibinfo {pages} {064610} (\bibinfo {year} {2006})},\
  \Eprint {http://arxiv.org/abs/hep-ph/0506186} {arXiv:hep-ph/0506186 [hep-ph]}
  \BibitemShut {NoStop}%
\bibitem [{\citenamefont {Songaila}\ and\ \citenamefont
  {Cowie}(2014)}]{Songaila}%
  \BibitemOpen
  \bibfield  {author} {\bibinfo {author} {\bibfnamefont {A.}~\bibnamefont
  {Songaila}}\ and\ \bibinfo {author} {\bibfnamefont {L.}~\bibnamefont
  {Cowie}},\ }\href {\doibase 10.1088/0004-637X/793/2/103} {\bibfield
  {journal} {\bibinfo  {journal} {Astrophys.J.}\ }\textbf {\bibinfo {volume}
  {793}},\ \bibinfo {pages} {103} (\bibinfo {year} {2014})},\ \Eprint
  {http://arxiv.org/abs/1406.3628} {arXiv:1406.3628 [astro-ph.CO]} \BibitemShut
  {NoStop}%
\bibitem [{\citenamefont {Molaro}\ \emph {et~al.}(2008)\citenamefont {Molaro},
  \citenamefont {Reimers}, \citenamefont {Agafonova},\ and\ \citenamefont
  {Levshakov}}]{alphaMolaro}%
  \BibitemOpen
  \bibfield  {author} {\bibinfo {author} {\bibfnamefont {P.}~\bibnamefont
  {Molaro}}, \bibinfo {author} {\bibfnamefont {D.}~\bibnamefont {Reimers}},
  \bibinfo {author} {\bibfnamefont {I.~I.}\ \bibnamefont {Agafonova}}, \ and\
  \bibinfo {author} {\bibfnamefont {S.~A.}\ \bibnamefont {Levshakov}},\ }\href
  {\doibase 10.1140/epjst/e2008-00818-4} {\bibfield  {journal} {\bibinfo
  {journal} {Eur.Phys.J.ST}\ }\textbf {\bibinfo {volume} {163}},\ \bibinfo
  {pages} {173} (\bibinfo {year} {2008})},\ \Eprint
  {http://arxiv.org/abs/0712.4380} {arXiv:0712.4380 [astro-ph]} \BibitemShut
  {NoStop}%
\bibitem [{\citenamefont {{Chand}}\ \emph {et~al.}(2006)\citenamefont
  {{Chand}}, \citenamefont {{Srianand}}, \citenamefont {{Petitjean}},
  \citenamefont {{Aracil}}, \citenamefont {{Quast}},\ and\ \citenamefont
  {{Reimers}}}]{alphaChand}%
  \BibitemOpen
  \bibfield  {author} {\bibinfo {author} {\bibfnamefont {H.}~\bibnamefont
  {{Chand}}}, \bibinfo {author} {\bibfnamefont {R.}~\bibnamefont {{Srianand}}},
  \bibinfo {author} {\bibfnamefont {P.}~\bibnamefont {{Petitjean}}}, \bibinfo
  {author} {\bibfnamefont {B.}~\bibnamefont {{Aracil}}}, \bibinfo {author}
  {\bibfnamefont {R.}~\bibnamefont {{Quast}}}, \ and\ \bibinfo {author}
  {\bibfnamefont {D.}~\bibnamefont {{Reimers}}},\ }\href {\doibase
  10.1051/0004-6361:20054584} {\bibfield  {journal} {\bibinfo  {journal}
  {Astron.Astrophys.}\ }\textbf {\bibinfo {volume} {451}},\ \bibinfo {pages}
  {45} (\bibinfo {year} {2006})},\ \Eprint
  {http://arxiv.org/abs/astro-ph/0601194} {astro-ph/0601194} \BibitemShut
  {NoStop}%
\bibitem [{\citenamefont {{Agafonova}}\ \emph {et~al.}(2011)\citenamefont
  {{Agafonova}}, \citenamefont {{Molaro}}, \citenamefont {{Levshakov}},\ and\
  \citenamefont {{Hou}}}]{alphaAgafonova}%
  \BibitemOpen
  \bibfield  {author} {\bibinfo {author} {\bibfnamefont {I.~I.}\ \bibnamefont
  {{Agafonova}}}, \bibinfo {author} {\bibfnamefont {P.}~\bibnamefont
  {{Molaro}}}, \bibinfo {author} {\bibfnamefont {S.~A.}\ \bibnamefont
  {{Levshakov}}}, \ and\ \bibinfo {author} {\bibfnamefont {J.~L.}\ \bibnamefont
  {{Hou}}},\ }\href {\doibase 10.1051/0004-6361/201016194} {\bibfield
  {journal} {\bibinfo  {journal} {Astron.Astrophys.}\ }\textbf {\bibinfo
  {volume} {529}},\ \bibinfo {eid} {A28} (\bibinfo {year} {2011})},\ \Eprint
  {http://arxiv.org/abs/1102.2967} {arXiv:1102.2967 [astro-ph.CO]} \BibitemShut
  {NoStop}%
\bibitem [{\citenamefont {Ade}\ \emph {et~al.}(2015)\citenamefont {Ade} \emph
  {et~al.}}]{Planck}%
  \BibitemOpen
  \bibfield  {author} {\bibinfo {author} {\bibfnamefont {P.~A.~R.}\
  \bibnamefont {Ade}} \emph {et~al.} (\bibinfo {collaboration} {Planck}),\
  }\href@noop {} {\  (\bibinfo {year} {2015})},\ \Eprint
  {http://arxiv.org/abs/1502.01589} {arXiv:1502.01589 [astro-ph.CO]}
  \BibitemShut {NoStop}%
\bibitem [{\citenamefont {Bagla}\ \emph {et~al.}(2003)\citenamefont {Bagla},
  \citenamefont {Jassal},\ and\ \citenamefont {Padmanabhan}}]{Bagla}%
  \BibitemOpen
  \bibfield  {author} {\bibinfo {author} {\bibfnamefont {J.~S.}\ \bibnamefont
  {Bagla}}, \bibinfo {author} {\bibfnamefont {H.~K.}\ \bibnamefont {Jassal}}, \
  and\ \bibinfo {author} {\bibfnamefont {T.}~\bibnamefont {Padmanabhan}},\
  }\href {\doibase 10.1103/PhysRevD.67.063504} {\bibfield  {journal} {\bibinfo
  {journal} {Phys. Rev.}\ }\textbf {\bibinfo {volume} {D67}},\ \bibinfo {pages}
  {063504} (\bibinfo {year} {2003})},\ \Eprint
  {http://arxiv.org/abs/astro-ph/0212198} {arXiv:astro-ph/0212198 [astro-ph]}
  \BibitemShut {NoStop}%
\bibitem [{\citenamefont {Padmanabhan}(2002)}]{Thanu}%
  \BibitemOpen
  \bibfield  {author} {\bibinfo {author} {\bibfnamefont {T.}~\bibnamefont
  {Padmanabhan}},\ }\href {\doibase 10.1103/PhysRevD.66.021301} {\bibfield
  {journal} {\bibinfo  {journal} {Phys. Rev.}\ }\textbf {\bibinfo {volume}
  {D66}},\ \bibinfo {pages} {021301} (\bibinfo {year} {2002})},\ \Eprint
  {http://arxiv.org/abs/hep-th/0204150} {arXiv:hep-th/0204150 [hep-th]}
  \BibitemShut {NoStop}%
\bibitem [{\citenamefont {Sandvik}\ \emph {et~al.}(2002)\citenamefont
  {Sandvik}, \citenamefont {Barrow},\ and\ \citenamefont {Magueijo}}]{BSBM}%
  \BibitemOpen
  \bibfield  {author} {\bibinfo {author} {\bibfnamefont {H.~B.}\ \bibnamefont
  {Sandvik}}, \bibinfo {author} {\bibfnamefont {J.~D.}\ \bibnamefont {Barrow}},
  \ and\ \bibinfo {author} {\bibfnamefont {J.}~\bibnamefont {Magueijo}},\
  }\href {\doibase 10.1103/PhysRevLett.88.031302} {\bibfield  {journal}
  {\bibinfo  {journal} {Phys. Rev. Lett.}\ }\textbf {\bibinfo {volume} {88}},\
  \bibinfo {pages} {031302} (\bibinfo {year} {2002})},\ \Eprint
  {http://arxiv.org/abs/astro-ph/0107512} {arXiv:astro-ph/0107512 [astro-ph]}
  \BibitemShut {NoStop}%
\end{thebibliography}%
\end{document}